\documentclass[12pt]{article}

\usepackage{epsfig}
\usepackage{cite}
\usepackage{amsmath, amssymb, amsfonts}
\usepackage{color}
\usepackage{latexsym}  
\usepackage{graphicx}
\usepackage{cancel}
\usepackage[colorlinks,bookmarks]{hyperref}
\hypersetup{pdfpagemode=UseNone, pdfstartview=FitH, linkcolor=blue, citecolor=red, urlcolor=blue}

\def\be{\begin{equation}}
\def\ee{\end{equation}}
\def\ba{\begin{eqnarray}}
\def\ea{\end{eqnarray}}

\def\bdm{\begin{displaymath}}
\def\edm{\end{displaymath}}

\def\bq{\begin{quote}}
\def\eq{\end{quote}}

 at 10truept

\newcommand{\bea}{\begin{eqnarray}}
\newcommand{\eea}{\end{eqnarray}}

\newcommand{\bi}{\begin{itemize}}
\newcommand{\ei}{\end{itemize}}

\newcommand{\beq}{\begin{equation}}
\newcommand{\eeq}{\end{equation}}
\newcommand{\beqa}{\begin{eqnarray}}
\newcommand{\eeqa}{\end{eqnarray}}

\def\12{{1 \over 2}}

\def\ltap{\ \raise.3ex\hbox{$<$\kern-.75em\lower1ex\hbox{$\sim$}}\ }
\def\gtap{\ \raise.3ex\hbox{$>$\kern-.75em\lower1ex\hbox{$\sim$}}\ }
\def\gl{\ \raise.5ex\hbox{$>$}\kern-.8em\lower.5ex\hbox{$<$}\ }
\def\roughly#1{\raise.3ex\hbox{$#1$\kern-.75em\lower1ex\hbox{$\sim$}}}

\begin{document}

\thispagestyle{empty}
\begin{flushright}
May 2026
\end{flushright}
\vspace*{1.5cm}
\begin{center}

{\Large \bf Chiral Electromagnetic Surface Waves on}
\vskip.3cm
{\Large \bf Chern--Simons Interfaces}

\vspace*{1.3cm} {\large
Nemanja Kaloper\footnote{\tt
kaloper@physics.ucdavis.edu} }\\
\vspace{.5cm}
{\em QMAP, Department of Physics and Astronomy, University of
California}\\
\vspace{.05cm}
{\em Davis, CA 95616, USA}\\

\vspace{1.5cm} ABSTRACT
\end{center}
We show that Maxwell theory with a codimension-$1$ Chern--Simons interface supports 
chiral electromagnetic surface waves on the interface, even when the bulk theory on 
both sides is conventional vacuum electrodynamics in infinite space. Solving the exact 
boundary value problem we find that the Chern--Simons interaction acts with opposite 
sign on the two helicities oriented along the interface, giving rise to one normalizable 
mode localized on the interface. This mode is a gapless chiral surface photon with linear 
dispersion and a frequency-independent index of refraction set by the Chern--Simons 
coefficient. This mode exists despite the absence of ambient material response or 
geometric confinement.

\vfill \setcounter{page}{0} \setcounter{footnote}{0}

\vspace{1cm}

\newpage

Topological terms localized to interfaces can induce qualitatively new phenomena in conventional bulk
theories. Chern--Simons couplings in electromagnetism provide
a particularly simple example. These terms do not modify local propagation in regions where
their coefficient is constant. However spatial variations of the coefficient introduce nontrivial boundary effects
that act directly on photon polarization and mode structure, as first emphasized in 
axion electrodynamics \cite{Wilczek:1987mv}. Such interface-borne terms arise in a
variety of effective field theories after integrating out heavy degrees of freedom, and are
most naturally understood as topological response phenomena whose physical consequences
are localized at interfaces \cite{Wen:1992ej}. Electromagnetic response effects 
associated with such $\theta$-terms and their
interfaces have been extensively discussed in the literature \cite{Essin:2008rq}. 
However a field theory of explicit propagating single-helicity electromagnetic modes,  
localized on the interface solely by the Chern--Simons terms, has not been identified. 
In this work, complementing those earlier studies of $\theta$-electrodynamics interfaces,
we show that the Chern--Simons interaction alone supports a
propagating, normalizable electromagnetic mode localized on the
interface. This mode arises from the
interplay of Maxwell field and the 
Chern--Simons boundary conditions, without any need for ambient material 
or additional propagating degrees of freedom.

Below we consider Maxwell theory in the presence of a codimension-$1$ Chern--Simons
interface. Our motivation comes from recalling that 
Chern--Simons terms are inherently chiral, coupling with opposite sign to different 
photon helicities at the interface. This acts as an impulsive momentum transfer 
whose sign flips between those two helicities. 
Thus, for bulk modes the interface boundary conditions are equivalent to a 
codimension-$1$ $\delta$-function potential, with an effective interaction that is
repulsive for one helicity and attractive for the other. This results in a unique normalizable, surface-localized 
mode for only one helicity parallel with the interface, whose localization weakens at low frequency. 
To look for such modes ``bound" to the interface,
we analyze the spectrum of the electromagnetic field focusing 
directly on field excitations supported at the Chern--Simons interface itself.

Starting from the ensuing equations of motion, we formulate the exact 
boundary value problem for electric and magnetic fields of 
the modes propagating along the
interface. Diagonalization in the $2D$ helicity eigenbasis along the interface reveals that one helicity
indeed experiences an attractive $\delta$-function interaction, confirming our intuition. 
Allowing physical momentum to be parallel to
the interface then produces a surface-bound photon with a 
definite helicity, and with dispersion relation fixed completely by the Chern--Simons
coefficient. This chiral surface photon realizes an intrinsic electromagnetic waveguide, 
in which confinement arises from topology rather than geometry or response of the ambient material.  

We couple Maxwell theory to a pseudo-scalar $\theta(x)$
through the Chern--Simons density,
\be
S = -\frac{1}{4}\int d^4x\, F_{\mu\nu}F^{\mu\nu}
- \frac{1}{4}\int d^4x\, \theta(x)\,
\epsilon^{\mu\nu\lambda\sigma} F_{\mu\nu}F_{\lambda\sigma} \, ,
\label{action}
\ee
which reduces to conventional electrodynamics in regions where $\theta$ is constant. 
In those regions the bulk photon propagation is not modified \cite{Deser:1981wh}. We
include the interface by specializing to configurations in which $\theta$ is 
piecewise constant, with a discontinuous jump across a codimension-$1$ hypersurface. Writing
$\theta(z) = \theta_- + \Delta\theta \, \Theta(z)$, 
the bulk term is a total derivative for $z\ne 0$, while the jump $\Delta\theta$ at $z=0$ 
induces a Chern--Simons interaction on the interface, a construction that appears
naturally in both field--theoretic and condensed matter realizations of
$\theta$-electrodynamics \cite{Qi:2008ew,Zanelli:2010zz}. We assume that the interface
is approximately planar on the scale of the electromagnetic wavelength. Thus 
translational invariance parallel to the surface is a good approximation, and propagation
along the interface is well defined. As a result we can identify the wall with the $z=0$ plane.

Treating $\theta(x)$ as a fixed given background
field, and varying the action with respect to the gauge field $A_\nu$ yields the
modified Maxwell equations
\be
\partial_\mu \Bigl(
F^{\mu\nu}
+ \theta(x)\,\epsilon^{\mu\nu\lambda\sigma} F_{\lambda\sigma}
\Bigr) = 0 \, ,
\label{eomF}
\ee
where external electric charges and currents vanish. Where $\theta$ is constant, 
electromagnetic dynamics reduces locally to ordinary
Maxwell theory, since the additional term in \eqref{eomF} is proportional to
$\epsilon^{\mu\nu\lambda\sigma}\partial_\mu F_{\lambda\sigma}$, which vanishes
identically as a consequence of the Bianchi identity
$\partial_{[\mu}F_{\lambda\sigma]}=0$ in Maxwell theory without magnetic
monopoles. Nontrivial effects of the Chern--Simons coupling arise only
from spacetime variations of $\theta(x)$. We consider configurations in which $\theta$ is piecewise constant,
with a discrete jump across the interface at $z=0$, 
\be
\theta(z) = \theta_- + \Delta\theta\,\Theta(z) \, ,
\label{thetawall}
\ee
where $\Theta(z)$ is the Heaviside step function. In this case
$\partial_z\theta = \Delta\theta\,\delta(z)$, and the Chern--Simons interaction
is localized on the interface, as claimed. 

To circumvent issues with gauge-fixing, we work directly with the field strengths $E^i = F^{0i}$ and
$B^i = \frac12 \epsilon^{ijk} F_{jk}$, which are 
automatically gauge-invariant and physical. The field equations 
\eqref{eomF} and the Bianchi identity $\partial_{[\mu}F_{\lambda\sigma]}=0$
then yield \cite{Kaloper:2026slg}
\ba
\vec \nabla\cdot\vec E &=& -\Delta\theta\,\delta(z)\,B_z  \, , \qquad \qquad \qquad
\vec \nabla\times\vec B - \partial_t\vec E
= \Delta\theta\,\delta(z)\,\hat z\times\vec E \, , \nonumber \\
\vec \nabla\cdot\vec B &=& 0 \, ,
\qquad \qquad \qquad \qquad \qquad  ~~\,
\vec \nabla\times\vec E + \partial_t\vec B = 0 \, ,
\label{fieldeq}
\ea
where we substituted the wall profile \eqref{thetawall}, and $\hat z$
is the unit vector normal to the interface. As a quick check we see that away 
from the interface Eqs. \eqref{fieldeq} reduce to the conventional vacuum Maxwell
equations. 

The physical effect of the Chern--Simons interaction on the interface is
enforced by the matching conditions encoded in 
Eqs. \eqref{fieldeq}. Integrating the scalar equations in
\eqref{fieldeq} over an infinitesimal pillbox whose faces are parallel to
the interface, and the vector equations around an infinitesimal rectangular
loop perpendicular to the interface, we find \cite{Kaloper:2026slg}
\ba
E_z(0^+) - E_z(0^-)
&=&
-\Delta\theta\,B_z(0) \, ,  \qquad ~ 
\hat z\times \bigl(\vec B_\parallel(0^+) - \vec B_\parallel(0^-) \bigr) =
\Delta\theta\,\hat z\times\vec E_\parallel(0) \, , \nonumber \\
B_z(0^+) - B_z(0^-)
&=& 0 \, ,
\qquad  \qquad \qquad ~ \,
\hat z\times \bigl(\vec E_\parallel(0^+) - \vec E_\parallel(0^-) \bigr) = 0 \, .
\label{BCs}
\ea
The first equation shows that the normal component of the electric field is
discontinuous at the wall whenever the magnetic field has a nonvanishing
normal component there. The second equation shows that the tangential
components of the magnetic field are also discontinuous when 
the tangential electric field does not vanish at the interface. The remaining two
equations enforce continuity of $B_z$ and of the tangential electric field. After 
these Gaussian pillbox integrations of \eqref{BCs} we find the complete set of boundary conditions
governing electromagnetic fields at a Chern--Simons interface. 

Since we are looking for solutions representing waves propagating along the interface and
localized to it, exponentially diminishing away in the normal direction, we take the ans\"atz
\be
\vec E = \vec{\cal E}(z)\,e^{i(\vec k_\parallel\cdot\vec x_\parallel-\omega t)} \, ,
\qquad \qquad
\vec B = \vec{\cal B}(z)\,e^{i(\vec k_\parallel\cdot\vec x_\parallel-\omega t)} \, ,
\label{waveans}
\ee
where $\vec x_\parallel=(x,y)$ are coordinates parallel to the wall and
$\vec k_\parallel$ is the conserved momentum along the interface.
Substituting into the bulk Eqs. \eqref{fieldeq} and combining
the results yields equations for the bulk wave profiles
on each side $(\pm)$, which have the same form
\be
\bigl(\partial_z^2 + \omega^2 - k_\parallel^2\bigr)\vec{\cal E}_\pm(z) = 0 \, ,
\qquad \qquad
\bigl(\partial_z^2 + \omega^2 - k_\parallel^2\bigr)\vec{\cal B}_\pm(z) = 0 \, ,
\label{wavebulk}
\ee
where $k_\parallel^2 = \vec k_\parallel^{\,2}$, and follow the same bulk dispersion relation
away from the interface, because the bulk on either side satisfies standard vacuum Maxwell equations. 
The electromagnetic fields on different sides of the 
interface are not the same, since the boundary conditions come 
from the Chern--Simons term. In particular, these fields are not invariant
under parity reflection across the interface. 

For $\kappa^2 =
k_\parallel^2-\omega^2 > 0$, the equations \eqref{waveans}
may yield exponentially decaying solutions
on both sides of the interface, if they simultaneously satisfy the boundary conditions 
\be
\vec{\cal E}_\pm(z) = \vec{\cal E}_\pm(0)\,e^{-\kappa|z|} \, ,
\qquad \qquad
\vec{\cal B}_\pm(z) = \vec{\cal B}_\pm(0)\,e^{-\kappa|z|} \, ,
\label{bulkprofile}
\ee
Those are the solutions we seek. We are thus isolating the surface-localized 
branch of the spectrum, distinct from the continuum of propagating bulk modes.

The integration constants $ \vec{\cal E}_\pm(0)$ 
and $ \vec{\cal B}_\pm(0)$ are not all independent. 
In the bulk, the firstÐorder Maxwell equations relate $\vec{\cal B}$ algebraically 
to $\vec{\cal E}$, while the divergence constraints relate the normal components 
of both fields to their tangential components. As a result, a complete set of independent 
integration constants may be chosen to be the tangential components of the electric field 
on each side of the interface. These amplitudes are fixed by the boundary 
conditions \eqref{BCs}, up to an overall normalization.

To proceed, away from the interface we use the ansatz
\eqref{waveans} which means that spatial derivatives acting on the bulk profiles reduce to 
$\vec \nabla \;\rightarrow\; i\vec k_\parallel \pm \kappa\,\hat z$, where the upper (lower) 
sign corresponds to the region $z>0$ ($z<0$). As above 
$\kappa^2 = k_\parallel^2-\omega^2>0$. The bulk Maxwell equations then yield 
linear algebraic relations between $\vec{\cal E}_\pm$ and
$\vec{\cal B}_\pm$. Specifically, Faraday's law $\vec \nabla\times\vec E + \partial_t\vec B = 0$ 
after straightforward algebra yields 
\be
\vec{\cal B}_\pm = \frac{1}{\omega} \Bigl(
\vec k_\parallel \times \vec{\cal E}_\pm \;\mp\;
i\kappa\,\hat z\times\vec{\cal E}_\pm \Bigr) \, .
\label{BfromE}
\ee
Similarly, Amp\`ereÕs law in vacuum $\vec \nabla\times\vec B - \partial_t\vec E = 0$ 
yields $ (i\vec k_\parallel \pm \kappa\,\hat z)\times\vec{\cal B}_\pm =
-\,i\omega\,\vec{\cal E}_\pm$. However, this equation is 
completely equivalent to \eqref{BfromE}, and so redundant, 
when the bulk on-shell condition $\omega^2 = k_\parallel^2-\kappa^2$ holds. Thus 
\eqref{BfromE} is the only independent relationship between integration constants that 
follows from vectorial Maxwell equations.

Additional relations come from the scalar Maxwell equations involving the divergence operator, 
which restrict the electric and magnetic field components normal to the interface. In the
absence of sources, $\vec \nabla\cdot\vec E=0$ and $\vec \nabla\cdot\vec B=0$ imply
\be
\vec k_\parallel\cdot\vec{\cal E}_\pm \;\mp\;
i\kappa\,{\cal E}_{z,\pm} = 0 \, ,
\qquad \qquad 
\vec k_\parallel\cdot\vec{\cal B}_\pm \;\mp\;
i\kappa\,{\cal B}_{z,\pm} = 0 \, .
\label{divconstraints}
\ee
Thus the normal components are fixed by
the longitudinal projections of the tangential components. Together with
Eq.~\eqref{BfromE}, these relations show that in each bulk region, an evanescent
mode is completely specified by the two tangential components of the electric
field.

The  boundary conditions \eqref{BCs} include the operator $\hat z\times$, 
which rotates tangential vectors. To disentangle those modes we 
switch to $2$-dimensional helicity eigenstates parallel 
to the interface since they are rotation eigenvectors, 
$\vec e_{\mathrm{L,R}} = (\hat x \pm i\hat y)/\sqrt{2}$, 
so that $\hat z\times \vec e_{\mathrm{L}} = -\,i\,\vec e_{\mathrm{L}}$ and 
$\hat z\times \vec e_{\mathrm{R}} = +\,i\,\vec e_{\mathrm{R}}$. 
Then for any vector $\vec V_\parallel = (V_x,V_y)$, the components in 
this basis are $V^{\mathrm{L,R}} =  \vec e^{~*}_{\mathrm{L,R}} \cdot \vec V$, or 
\be
V^{\mathrm{L}} \equiv \frac{1}{\sqrt{2}}\,(V_x - i V_y) \, , 
\qquad \qquad
V^{\mathrm{R}} \equiv \frac{1}{\sqrt{2}}\,(V_x + i V_y) \, .
\label{circdef}
\ee
The last line of Eq. \eqref{BCs}, obtained by integrating Faraday's law 
across the interface, implies that the normal component of the
magnetic field and the tangential component of the electric field are continuous
on the interface: 
${\cal B}_{z,+}(0) = {\cal B}_{z,-}(0)$ and $\vec {\cal E}_{\parallel, +}(0) = \vec {\cal E}_{\parallel,-}(0)$. 
In the $2D$ helicity eigenbasis,  
\be
{\cal B}_{z,+}(0) = {\cal B}_{z,-}(0)  = {\cal B}_{z}(0) \, \, , 
\quad {\cal E}^{\mathrm{L}}_+(0) = {\cal E}^{\mathrm{L}}_-(0) = {\cal E}^{\mathrm{L}}(0) \, ,
\quad
{\cal E}^{\mathrm{R}}_+(0) = {\cal E}^{\mathrm{R}}_-(0) = {\cal E}^{\mathrm{R}}(0) \, .
\label{Econt}
\ee

The tangential boundary condition in \eqref{BCs} is manifestly chiral since 
it includes the rotation generator $\hat z \times$. Expanding the vector fields 
in the $2D$ helicity eigenstate basis on each side of the interface and bearing in mind 
that $\vec {\cal E}_{\parallel}$ is continuous on the interface then yields 
\be
{\cal B}^{\mathrm{L}}_+(0) - {\cal B}^{\mathrm{L}}_-(0) =
\Delta\theta\,{\cal E}^{\mathrm{L}}(0) \, ,
\qquad \qquad
{\cal B}^{\mathrm{R}}_+(0) - {\cal B}^{\mathrm{R}}_-(0) =
\Delta\theta\,{\cal E}^{\mathrm{R}}(0) \, .
\label{BCcirc}
\ee
The final boundary condition, the first Eq. in \eqref{BCs}, involves only the 
normal components of the electric and magnetic fields, 
and hence simply carries over to the helicity eigenbasis: 
\be
{\cal E}_{z,+}(0) - {\cal E}_{z,-}(0) = -\Delta\theta\,{\cal B}_{z}(0) \, .
\label{normbc}
\ee
This completes the reformulation of the boundary conditions in the helicity eigenbasis.

Finally we combine the boundary conditions in Eqs. \eqref{Econt}, \eqref{BCcirc} and \eqref{normbc}
and the relations we obtained from bulk Maxwell equations. Away from 
the interface, the bulk solutions satisfy \eqref{BfromE}.
Projecting that equation onto the interface-parallel helicity eigenstates,
\be
{\cal B}^{\mathrm{L}}_\pm(0) = \frac{1}{\omega}
\Bigl[ (\vec k_\parallel \times \vec{\cal E})^{\mathrm{L}}
\;\pm\; \kappa\,{\cal E}^{\mathrm{L}}(0) \Bigr] \, , 
\qquad \qquad
{\cal B}^{\mathrm{R}}_\pm(0) = \frac{1}{\omega}
\Bigl[ (\vec k_\parallel \times \vec{\cal E})^{\mathrm{R}} 
\;\mp\; \kappa\,{\cal E}^{\mathrm{R}}(0) \Bigr] \, .
\label{Bproj}
\ee
Here we have used 
$\vec{\cal E}^{\mathrm{L,R}}_+(0)=\vec{\cal E}^{\mathrm{L,R}}_-(0)$, and 
$\hat z\times\vec{\cal E}^{\mathrm{L}}=-i\vec{\cal E}^{\mathrm{L}}$, 
$\hat z\times\vec{\cal E}^{\mathrm{R}}=+i\vec{\cal E}^{\mathrm{R}}$, which follow 
from the continuity of components of $\vec {\cal E}_{\parallel}$ 
in the helicity eigenbasis. Eq. \eqref{Econt}. 

Next, the terms
$\propto \vec k_\parallel\times\vec{\cal E}$ cancel in the difference 
of the magnetic fields across the interface: 
\be
{\cal B}^{\mathrm{L}}_+(0) - {\cal B}^{\mathrm{L}}_-(0) =
\frac{2\kappa}{\omega}\,{\cal E}^{\mathrm{L}}(0) \, , 
\qquad \qquad
{\cal B}^{\mathrm{R}}_+(0) - {\cal B}^{\mathrm{R}}_-(0) =
-\frac{2\kappa}{\omega}\,{\cal E}^{\mathrm{R}}(0) \, .
\label{Bjump}
\ee
Substituting Eqs. \eqref{Bjump} into the helicity eigenbasis 
boundary conditions \eqref{BCcirc} yields 
\be
\frac{2\kappa}{\omega}\,{\cal E}^{\mathrm{L}}(0) =
\Delta\theta\,{\cal E}^{\mathrm{L}}(0) \, , 
\qquad \qquad
-\frac{2\kappa}{\omega}\,{\cal E}^{\mathrm{R}}(0) =
\Delta\theta\,{\cal E}^{\mathrm{R}}(0) \, .
\label{existcond}
\ee
These equations provide the crucial information controlling the existence of the solutions
which satisfy our initial ans\"atz. 
For nontrivial solutions ${\cal E}^{\mathrm{L,R}}(0) \ne 0$ these equations require
\be
\kappa = \frac{\Delta\theta\,\omega}{2}
\quad \text{(left-handed)} \, ,
\qquad \text{or} \qquad
\kappa = -\frac{\Delta\theta\,\omega}{2}
\quad \text{(right-handed)} \, .
\label{kappacond}
\ee
Normalizability of the evanescent bulk profiles requires $\kappa>0$, so for
$\omega>0$ exactly one helicity sector admits a localized solution, depending
on the sign of $\Delta\theta$. Without any loss of generality we can take
$\Delta\theta>0$, which selects the left-handed polarization 
for the localized mode. For $\Delta\theta<0$, the roles of the two helicities 
are reversed. Importantly, \eqref{kappacond} reveal that the ``decay constant"
$\kappa$ is not a mass, since $\kappa \propto \omega$. Hence 
the spectrum of the localized mode is gapless. 
Note that \eqref{kappacond} implies that after we fix the integration constants to solve
the boundary conditions for the selected helicity, no additional normalizable solutions exist.

To complete the explicit construction of the localized mode, we can therefore focus 
on the left-handed modes. The tangential electric field continuity \eqref{Econt} yields 
${\cal E}^{\mathrm{L}}_+(0)={\cal E}^{\mathrm{L}}_-(0)\equiv{\cal E}_0$, where
${\cal E}_0$ is an arbitrary constant, and so 
\be
\vec{\cal E}_{\parallel,\pm}(z) =
{\cal E}_0\,\vec e_{\mathrm{L}}\,e^{-\kappa|z|} \, .
\label{Etanprofile}
\ee
The normal component of the electric field is fixed by Gauss' law in the bulk,
$\vec\nabla\cdot\vec E=0$, which implies
$\vec k_\parallel\cdot\vec{\cal E}_\pm\mp i\kappa{\cal E}_{z,\pm}=0$. Solving
for ${\cal E}_{z,\pm}$ yields
\be
{\cal E}_{z,+}(0) = \frac{i}{\kappa}\,
(\vec k_\parallel\cdot\vec e_{\mathrm{L}})\, {\cal E}_0 \, ,
\qquad \qquad 
{\cal E}_{z,-}(0) = -\frac{i}{\kappa}\,
(\vec k_\parallel\cdot\vec e_{\mathrm{L}})\, {\cal E}_0 \, .
\label{Ezvalues}
\ee
Thus the normal component of the electric field is odd across the interface, as we noted above; 
it jumps by ${\cal E}_{z,+}(0) - {\cal E}_{z,-}(0) = \frac{2i}{\kappa} (\vec k_\parallel\cdot
\vec e_{\mathrm{L}})\,{\cal E}_0$. To find ${\cal B}_{z,\pm}(0)$ 
we substitute the solution $\vec{\cal E}_\pm(0) = {\cal E}_0\,\vec e_{\mathrm{L}}$ into 
Eq. \eqref{BfromE}, and bear in mind that $(\hat z\times \vec e_{\mathrm{L}})_z = 0$ -- so that
we can drop the subscripts $\pm$. We find  
${\cal B}_{z}(0) = \frac{{\cal E}_0}{\omega} (\vec k_\parallel \times \vec e_{\mathrm{L}})_z$. 
Since both $\vec k_\parallel$ and $\vec e_{\mathrm{L}}$ lie in the interface plane, 
$(\vec k_\parallel \times \vec e_{\mathrm{L}})_z
= -\,i\,(\vec k_\parallel\cdot\vec e_{\mathrm{L}})$, so that  
$ {\cal B}_z(0) = -\,\frac{i}{\omega}\, (\vec k_\parallel\cdot\vec e_{\mathrm{L}})\,{\cal E}_0$. 
Substituting into \eqref{normbc} yields
\be
-\Delta\theta\,{\cal B}_z(0)
= \frac{i\,\Delta\theta}{\omega}\,
(\vec k_\parallel\cdot\vec e_{\mathrm{L}})\,{\cal E}_0 \, ,
\ee
which independently matches the jump in ${\cal E}_z$ provided
$\kappa = \frac{\Delta\theta\,\omega}{2}$.

To get the wave profiles for both solutions -- the localized one and the 
non-normalizable counterpart -- we combine the tangential and normal 
field strengths. For the localized solutions, the electric field is
\be
\vec{\cal E}_\pm(z) = {\cal E}_0
\left[ \vec e_{\mathrm{L}} \;\pm\;
\frac{i}{\kappa}\, (\vec k_\parallel\cdot\vec e_{\mathrm{L}})\,
\hat z \right] e^{-\kappa|z|} \, .
\label{Efinal}
\ee
The magnetic field follows from Faraday's law
\eqref{BfromE}. Substituting \eqref{Efinal} into \eqref{BfromE} gives
\be
\vec{\cal B}_\pm(z) =
\frac{{\cal E}_0}{\omega} \left[
\vec k_\parallel\times\vec e_{\mathrm{L}}
\;\pm\; \kappa\,\vec e_{\mathrm{L}} \right] e^{-\kappa|z|} \, .
\label{Bfinal}
\ee
The term $\vec k_\parallel\times\vec e_{\mathrm{L}}$ is normal to the
interface and provides the entire $B_z$ component, while the term proportional
to $\vec e_{\mathrm{L}}$ is tangential. The overall normalization ${\cal E}_0$
remains arbitrary, to be set by the initial conditions. For the localized mode, 
$\kappa > 0$ \eqref{kappacond}; the other mode corresponds to 
$\mathrm{L} \rightarrow \mathrm{R}$ and the associated sign flip of $\kappa$. Upon $\omega \leftrightarrow -\omega$,
these solutions flip around. 

The wavefunction ``decay constant" $\kappa$ is not 
an independent parameter. From the ans\"atz \eqref{bulkprofile}, and \eqref{kappacond}, 
\be
\kappa^2 = k_\parallel^2 - \omega^2  \, , 
\qquad \qquad  \kappa^2 = \frac{(\Delta\theta)^2\,\omega^2}{4} \, ,
\label{kappadef}
\ee
where $k_\parallel = |\vec k_\parallel|$ is the conserved momentum along the
interface. Combining these,
\be
\omega^2 =
\frac{k_\parallel^2}{1 + \frac{\Delta\theta^{\,2}}{4}} \, .
\label{disp2}
\ee
We note that our normalizable mode can be identified with a pole in the electromagnetic reflection 
coefficient at the interface. That confirms that it corresponds to a genuine propagating excitation of the system.

The dispersion relation \eqref{disp2} is unexpected -- given the context where it appears -- in that 
it is linear at all scales below the UV cutoff, 
with a coefficient smaller than unity and independent of scale. As a
result,  both the phase velocity and the group velocity of the surface photon are
\be 
v_{\rm ph} = v_{\rm g}
= \frac{1}{\sqrt{1+\frac{1}{4}(\Delta\theta)^2}} < 1 \, ,
\label{velos}
\ee
and so are strictly less than the speed of light in the vacuum bulk, and independent of
frequency. The reduction in propagation speed is set by the
Chern--Simons jump $\Delta\theta$ and does not depend on any microscopic details
of the interface. This result is robust all the way
to the cutoff of the effective field theory, where the structure of the interface is microscopically
resolved. This roughly occurs at the scale given by the inverse thickness of the interface. 

Below this cutoff, the localized mode propagates as light in a homogeneous medium
with a constant refractive index $n_{\rm eff}=\sqrt{1+\tfrac{1}{4}(\Delta\theta)^2}$. 
It is nondispersive, in the sense that $v_{\rm ph} = v_{\rm g}$, although 
the bulk on either side of the
interface is the vacuum without any ambient material response present. The Chern--Simons
interface behaves as an infinitesimally thin, frequency-independent optical
medium whose refractive properties are fixed by topology alone. It is  
effectively trapping an electromagnetic wave without any assistance by ambient material 
degrees of freedom. Note that since the ``decay constant" 
$\kappa \propto \omega$, Eq. \eqref{kappadef}, the 
lowest frequencies are the least localized ones.

This behavior sharply distinguishes our Chern--Simons waveguide from conventional
ones. In systems such as parallel conducting
plates, transverse confinement introduces a frequency gap, and at high frequency 
the modes behave as free photons in the bulk 
so that confinement effectively ceases \cite{Jackson:1998nia}. By contrast, the
surface mode supported by a Chern--Simons interface is gapless and remains slow
at all frequencies below the cutoff. There is no regime in which it merges with a bulk photon or
forgets that it is interface supported. More generally, 
different $\Delta\theta$ may label distinct topological sectors of the interface,
each supporting its own chiral surface photon with a characteristic propagation
speed and localization length. 

In addition, our localized mode is not equivalent to the photon of
intrinsic $(2+1)$-dimensional Maxwell theory \cite{Deser:1981wh}, 
although it propagates a single degree of freedom on 
a $(2+1)$-dimensional interface. The intrinsic $2+1$ Maxwell mode is parity and
time-reversal invariant and is dual to a scalar field. In contrast our 
interface mode is manifestly chiral, with its polarization fixed by the
sign of the Chern--Simons discontinuity. 
As a result, the theory of the chiral interface mode is not invariant under time reversal,
since it maps the allowed excitation into a mode of opposite helicity, which is 
not supported by the interface. 

Finally, as a check of the efficiency of mode localization 
to the interface, we compute the Poynting vector of the localized wave.
Since the solution is monochromatic, it suffices to consider the
time-averaged energy flux, which for harmonic fields
$\vec E = \Re\!\left[\vec{\cal E}(z)\,
e^{i(\vec k_\parallel\cdot\vec x_\parallel-\omega t)}\right]$ and 
$\vec B = \Re\!\left[\vec{\cal B}(z)\,e^{i(\vec k_\parallel\cdot\vec x_\parallel-\omega t)}\right]$
is given by
\be
\langle \vec S \rangle = \frac{1}{2\mu_0}\,
\Re\!\left(\vec{\cal E}\times\vec{\cal B}^{\,*}\right) \, .
\label{PoyntingDef}
\ee
Substituting the explicit localized solution obtained in Eqs. \eqref{Efinal} and \eqref{Bfinal}
shows that all contributions are proportional to $e^{-2\kappa|z|}$, reflecting the
exponential localization of the energy flux near the interface. The normal component
vanishes, $ \langle S_z \rangle = 0$, and the energy 
flux is purely tangential, pointing along $\vec k_\parallel$:
\be
\langle \vec S(z) \rangle = \frac{|{\cal E}_0|^2}{2\mu_0\,\omega}\,
\vec k_\parallel\, e^{-2\kappa|z|} \, .
\label{PoyntingFinal}
\ee
Equation \eqref{PoyntingFinal} provides a complete description of the energy
transport carried by the surface mode. Energy density is manifestly non-negative by 
construction. The absence of any normal component of the
Poynting vector confirms that the mode is bound to the interface and does not radiate
into the surrounding bulk.

We can recap our results here. We have shown that Maxwell electrodynamics with a codimension-$1$
Chern--Simons interface necessarily supports an interface-localized electromagnetic
mode of definite helicity. This mode does not rely on any
modification of bulk photon dynamics. 
Instead, it comes about because the interface-borne electromagnetic Chern--Simons 
interaction acts with opposite sign on the two photon helicities. 
As a result, one polarization experiences an attractive $\delta$-function 
interaction and forms a localized state on the interface.  
This surface mode is therefore strictly chiral, with its handedness
fixed by the sign of the Chern--Simons jump $\Delta\theta$. While the 
resulting mode carries a single propagating degree of freedom, 
it is not equivalent to Maxwell theory intrinsic to $(2+1)$ dimensions. Our 
interface-localized excitation is fundamentally chiral and so not time-reversal invariant, 
reflecting helicity selection enforced by its $4$-dimensional origin.

The localized chiral photon propagates with a universal, linear dispersion
relation, featuring a constant index of refraction. The index of refraction is fixed 
by $\Delta\theta$ for all momenta below the cutoff of the interface effective field theory. 
Both the phase and group velocities are reduced relative to the speed of light in the bulk. 
We find it remarkable that the propagation of the surface photon is identical in form to that of
light traveling in a homogeneous medium with constant refractive index, even
though the bulk on either side of the interface is vacuum and no ambient material degrees
of freedom are present. In this sense the dispersion relation \eqref{disp2} is structural rather
than dynamical, closely paralleling the geometric origin of the Pancharatnam phase
in polarization rotation across Chern--Simons interfaces \cite{Kaloper:2026ygk}.
This persists all the way up to the ultraviolet cutoff of the effective theory,
set roughly by the inverse thickness of the interface, separating two different $\theta$-vacua. 
Clearly, near the cutoff the corrections will appear, and force a more detailed description 
that requires a UV completion of the theory \eqref{action}. 
 
In closing, our results demonstrate that Maxwell theory admits intrinsic chiral surface  
excitations when supplemented with a Chern--Simons interface. The surface photon is
not a modified bulk mode nor a medium-induced fluctuation. Instead, it is 
a boundary degree of freedom supported by the interface topology. Thus our results 
point at a novel regime of electromagnetic boundary physics,  
and provides a new insight on the interplay between topology and electromagnetic field dynamics.

\end{document}